\documentstyle[12pt,fps]{article}
\newcommand{\Z}{{\sf Z \!\!\! Z}}

\makeatletter
\@addtoreset{equation}{section}
\makeatother
 
\title{Perfect Actions with Chemical Potential
\footnote{This work is supported in part by funds provided by the U.S.
Department of Energy (D.O.E.) under cooperative research agreement
DE-FC02-94ER40818.}}
\author{W. Bietenholz$^{{\rm a}}$ and 
U.-J. Wiese$^{{\rm b}}$ \\ \\
$^{{\rm a}}$ HLRZ c/o Forschungszentrum J\"{u}lich \\
52425 J\"{u}lich, Germany \\ \\
$^{{\rm b}}$
Center for Theoretical Physics, \\
Laboratory for Nuclear Science, and Department of Physics \\
Massachusetts Institute of Technology (MIT) \\
Cambridge, Massachusetts 02139, U.S.A. \\ \\
Preprint HLRZ 1997-71 \\ \\}
 
\begin{document}
\maketitle
\begin{abstract} \normalsize

We show how to include a chemical potential $\mu$ in perfect lattice 
actions. It turns out that the standard procedure of multiplying the 
quark fields $\Psi, \bar\Psi$ at Euclidean
time $t$ by $\exp(\pm \mu t)$,
respectively, is perfect. As an example, the case of free 
fermions with 
chemical potential is worked out explicitly. Even after truncation,
cut-off effects in the pressure and the baryon density
are small. Using a (quasi-)perfect action, numerical QCD simulations
for non-zero chemical potential 
become more powerful, because coarse lattices are sufficient for 
extracting continuum physics.

\end{abstract}
 
\maketitle
 
\newpage

Understanding strongly interacting matter at {\em finite baryon
density} is a long-standing and challenging problem,
motivated for instance by relativistic heavy ion collision
and by the physics of neutron stars.
The standard procedure to formulate lattice QCD at a finite 
chemical potential $\mu$ includes a factor $\exp (\pm \mu )$ 
in the time-like link variables \cite{HasKar}.
As a consequence, the Euclidean action is complex, the Boltzmann 
factor cannot be interpreted as a probability, and standard 
Monte Carlo techniques fail. 

The usual method to handle a chemical potential is to simulate
at $\mu = 0$, and to include the baryon number term by some
re-weighting technique in measured observables \cite{Barb90}.
However, this method is tractable only on small physical volumes
$V$, for a recent review see Ref.~\cite{Japan}. 
The essential numerical problem is to measure
exponentially suppressed observables, like the partition function
ratio $Z(\mu )/Z(0) \sim \exp (-\beta V [f(\mu )-f(0)])$.
Here $\beta$ is the inverse temperature and $f(\mu )$ is the
free energy density. In numerical simulations the above ratio
arises as an average over many positive and negative contributions.
Hence its accurate determination requires tremendous statistics.
An improved lattice action can not directly solve this sign problem,
but it would help because it suppresses the artifacts due to the
finite lattice spacing.


As a particularly troublesome effect caused by lattice
artifacts, there is an upper limit for the possible fermion
number density on the lattice.
The value of this limit depends on the lattice action.
It can be understood from the maximal coupling
distance in the time-like direction, which restricts the
exponential growth of the leading term at $\mu \to \infty$.
An improved action can weaken this unphysical saturation effect, 
and push it to a larger chemical potential.

As a further problem, chiral symmetry is already restored 
--- and a non-zero quark number sets in ---
as $\mu$ reaches half the pion mass \cite{barb}.
Generally the quenched approximation has been blamed for that
threshold \cite{Gibbs}, but --- at least with respect to the quark
number --- it persists in simulations using the 
Glasgow method \cite{Glas}.
It could be a further manifestation of lattice artifacts,
which might cause a spurious scattering of the eigenvalues \cite{Vink}.
This problem further motivates the use of improved lattice actions
at finite baryon density, since they render the observables more
continuum like than a standard lattice formulation.

{\em Perfect lattice actions} provide a possibility to 
completely eliminate cut-off
effects. In practice, approximately perfect actions have been
constructed, and they have led to a significant reduction of finite
lattice spacing artifacts \cite{Has94,DeG95,Bie96,LAT96}. 
When applied to a 
perfect action at $\mu = 0$, the standard procedure of multiplying the
fermion fields $\Psi, \bar\Psi$ at time $t$ by $\exp(\pm \mu t)$,
respectively, turns out to be quantum perfect in the fully interacting
theory. We use the example of a free
quark to demonstrate that --- even after truncation --- the resulting
lattice action approximates continuum physics very well. 
\footnote{Recently another paper has appeared, where a chemical
potential was included in a (classically) perfect action
of the 2d $O(3)$ model \cite{O3}.} On coarse 
lattices, the scaling behavior of observables is strongly improved 
compared to the standard action and other proposals.
For discussions of the free standard lattice fermions at $\mu >0$,
see the second Ref. in~\cite{HasKar} and Refs.~\cite{barb,MaSto}.

Let us consider a block factor $n$ renormalization group 
transformation, which maps a fine lattice theory of link variables
$U$ and fermion fields $\Psi, \bar\Psi$ onto a coarse lattice theory
with fields $U', \Psi', \bar\Psi'$. The actions 
$S[U,\Psi,\bar\Psi]$ and $S'[U',\Psi',\bar\Psi']$ on the fine and
coarse lattice at $\mu = 0$ are related by
\begin{equation}
\exp(-S'[U',\Psi',\bar\Psi'])=\int {\cal D}U {\cal D} \Psi 
{\cal D} \bar\Psi \exp \Big( -S[U,\Psi,\bar\Psi]
-T[U',\Psi',\bar\Psi',U,\Psi,\bar\Psi] \Big) \ ,
\end{equation}
where $T$ characterizes the renormalization group transformation.
It must be chosen such that the partition function remains
invariant, $Z' = Z$.
Starting on a very fine lattice with quark mass $m/(n N)$ and
inverse temperature $\beta n N$, and performing $N$ renormalization
group steps with block factor $n$, leads in the limit $n N \rightarrow
\infty$ to a perfect action $S^*[U,\Psi,\bar\Psi]$ for quarks
of mass $m$ at inverse temperature $\beta$. 

Now we address the question how to incorporate the chemical 
potential. In the continuum this can be done simply by a
substitution in momentum space,
\begin{equation} \label{sub}
\Psi (k) \to \ ^{\mu}\Psi (\vec k , k_{4} - i \mu ) \ , \
\bar \Psi (k) \to \ ^{\mu}\bar \Psi (\vec k ,k_{4}-i\mu) .
\end{equation}
In perturbation theory, it is equivalent to leaving the fermionic fields
unchanged and replacing 
\begin{equation} \label{subs}
k_{4} \to k_{4}+i\mu
\end{equation}
in the free propagator instead.
A priori it is not obvious how to transfer this substitution to
a standard lattice formulation. 
An early guess replaced $\sin k_{4} \to \sin k_{4} + i \mu$ 
in the fermionic propagator, but this
formulation does not have a proper continuum limit. Instead
it turned out that one should use exactly the same substitution
(\ref{sub}) resp. (\ref{subs}) also on the lattice to incorporate
$\mu$ in the actions resp.
propagators of Wilson or staggered fermions \cite{HasKar}.

Hence, on the initial fine lattice a chemical potential $\mu$ can be
introduced by replacing $\Psi(\vec x,t)$ and $\bar \Psi(\vec x,t)$ 
($t \in [0,\beta]$) in the Wilson action $S[U,\Psi,\bar\Psi]$ by
\begin{equation}
\label{mupsi}
^\mu\Psi(\vec x,t) = \exp(\mu t) \Psi(\vec x,t), \
^\mu\bar\Psi(\vec x,t) = \exp(- \mu t) \bar\Psi(\vec x,t).
\end{equation}
The chemical potential appears as a purely imaginary constant Abelian
gauge potential $A_4 = i \mu$. Thus, the fields $^\mu\Psi(\vec x,t)$ and
$^\mu\bar\Psi(\vec x,t)$ can be viewed as being parallel
transported to $t=0$. This ensures covariance under the corresponding
Abelian gauge transformations. In the presence of a chemical 
potential, the renormalization group transformation must be modified
if one wants to manifestly preserve this additional symmetry. 
This is achieved naturally by using 
$T[U',^{\mu n}\Psi',^{\mu n}\bar\Psi',U,^\mu\Psi,^\mu\bar\Psi]$ with
\begin{equation}
^{\mu n}\Psi'(\vec x',t') = \exp(\mu n t') \Psi'(\vec x',t'), \
^{\mu n}\bar\Psi'(\vec x',t') = \exp(- \mu n t') 
\bar\Psi'(\vec x',t').
\end{equation}
Note that on the coarse lattice $t' \in [0,\beta/n]$. Performing the
corresponding renormali-zation group transformation results in the
action $S'[U',^{\mu n}\Psi',^{\mu n}\bar\Psi']$. Indeed, this is
exactly what one obtains by applying the standard procedure to
include a chemical potential $\mu n$ directly in the coarse lattice
action $S'$. Starting with a small chemical potential $\mu/(n N)$,
and again iterating the blocking procedure $N$ times, yields the
perfect action $S^*[U,^\mu\Psi,^\mu\bar\Psi]$ with chemical potential
$\mu$ in the limit $n N \rightarrow \infty$. Hence, performing the
standard procedure of including $\mu$ in a perfect action --- 
constructed at $\mu=0$ --- yields a perfect action at arbitrary $\mu$.
We emphasize that this argument applies to the fully interacting
quantum theory.

As a special case, in perturbation theory 
one may send the blocking factor $n \to
\infty$. Then we don't need to iterate any more; the perfect
action is obtained for $N=1$. We call this technique ``blocking from
the continuum'' \cite{Bie96}, because one does not start the blocking
process from a fine lattice but directly from the continuum theory.
From that procedure it is particularly evident that the continuum
relation (\ref{subs}) is inherited without alteration by the
perfect lattice propagator. This agrees with the result that we derived 
above using a finite blocking factor.

For a free fermion with mass $m$ at $\mu = 0$, 
a perfect action has been constructed
in Ref. \cite{Bie96},
\begin{eqnarray}
\label{perfprop}
&&S^*[\Psi,\bar\Psi] = \int_B \frac{d^{4}k}{(2 \pi)^4} \
\bar\Psi(-k) \Delta^{-1}(k) \Psi(k), \nonumber \\
&&\Delta(k) = \sum_{l \in \Z^4} 
\frac{\Pi^{2}(k + 2 \pi l)}{i \gamma_\mu (k_\mu + 2 \pi l_\mu) + m} +
\frac{1}{\alpha}, \nonumber \\
&&\Pi(k) = \prod_\mu \frac{\hat k_\mu}{k_\mu}\ , \quad \hat k_\mu =
2 \sin \frac{k_\mu}{2}.
\end{eqnarray}
Here, $B=]-\pi ,\pi ]^{4}$ is the Brillouin zone and
$\alpha$ is a parameter in the renormalization group 
transformation. At $\alpha = \infty$ and $m=0$, 
the perfect action is chirally invariant, and 
hence --- in agreement with the Nielsen-Ninomiya theorem --- the 
action is nonlocal \cite{Wie93}. 
At finite $\alpha$, on the other hand, chiral symmetry
is explicitly broken in the action --- though still present
in the observables \cite{Schwing} --- and the
action becomes local. In particular, for $\alpha = (e^m - m - 1)/m^2$, 
locality --- in the sense of an exponential decay of the couplings ---
is optimal.

Applying the standard method of including the chemical potential, the
perfect action at $\mu>0$ takes the form $S^*[^\mu\Psi,^\mu\bar\Psi]$.
In momentum space
we obtain
\begin{equation}
S^*[^\mu\Psi,^\mu\bar\Psi] = \int_B \frac{d^{4}k}{(2\pi )^{4}} \
\bar\Psi(-k) \Delta^{-1}(\vec k,k_4 + i \mu) \Psi(k).
\end{equation}

Let us consider the cut-off effects in the pressure $p$ and the
baryon number density $n_{B}$ on the lattice. In the continuum, 
for massless quarks at zero temperature, these quantities are 
given by
\begin{equation}
p = \frac{\mu^4}{6 \pi^2} \quad , 
\quad n_{B} = \frac{2 \mu^3}{9 \pi^2} \ .
\end{equation}
On the lattice they take the form
\begin{eqnarray}
p &=& \int_B \frac{d^{4}k}{(2 \pi)^4} \
\Big[ \log det \Delta(\vec k,k_4) - \log det \Delta(\vec k,k_4 + i \mu)
\Big] , \nonumber \\
n_{B} &=& \frac{1}{3} \frac{\partial}{\partial \mu} p\ .
\end{eqnarray}
The lattice propagator for a perfect action is given in 
eq.~(\ref{perfprop}), but one may also insert other fermionic
lattice propagators known from the literature, such as the
Wilson fermion, staggered fermions or the so-called D234 action
\cite{D234}.
The latter is an extension of the Wilson fermion in the spirit
of Symanzik's improvement program: the
artifacts due to the finite lattice spacing are
cancelled in the leading order
by additional couplings along the axes.

In order to make the perfect action applicable, its couplings have
to be truncated to a short range. Hence a fair comparison to other
lattice fermions can only deal with a truncated perfect fermion.
In Ref.~\cite{LAT96} we performed such a truncation for the perfect
fermion action (\ref{perfprop}) --- with the parameter $\alpha$
optimizing locality --- to the couplings in a unit hypercube. 
Figure 1 illustrates the cut-off effects in the ratio
$p/\mu^4$ as a function
of $\mu$ in lattice units for the zoo of lattice fermions
mentioned above.
In the continuum limit $\mu \rightarrow 0$, 
they all approach the correct value $1/6 \pi^2$. At finite $\mu$, 
the Wilson action suffers from a severe cut-off dependence. 
The behavior is somewhat better for staggered fermions, but the 
artifacts are still quite bad. The D234 action is successful at small
$\mu$, but around $\mu = 1.5$ it collapses completely. The action of the 
``hypercube fermion'' (truncated perfect fermion), however,
remains close to the continuum value of the considered ratio
over a wide range of chemical potentials.
For example, the considered ratio deviates by less than $50 \%$ from
the continuum value up to $\mu=0.8$ for the Wilson and the staggered 
fermion, up to $\mu=1.6$ for the D234 action, and up to $\mu =8.2$
for the hypercube fermion. We note that an alternative truncated
fixed point fermion, which has been
constructed by the Bern/Boulder group \cite{FN}, shows a similar 
scaling quality; it stays even a little closer to the continuum value
than the ``hypercube fermion'' in Figure 1.


\begin{figure}[hbt]
\def\fpsangle{270}
\epsfxsize=100mm
\fpsbox{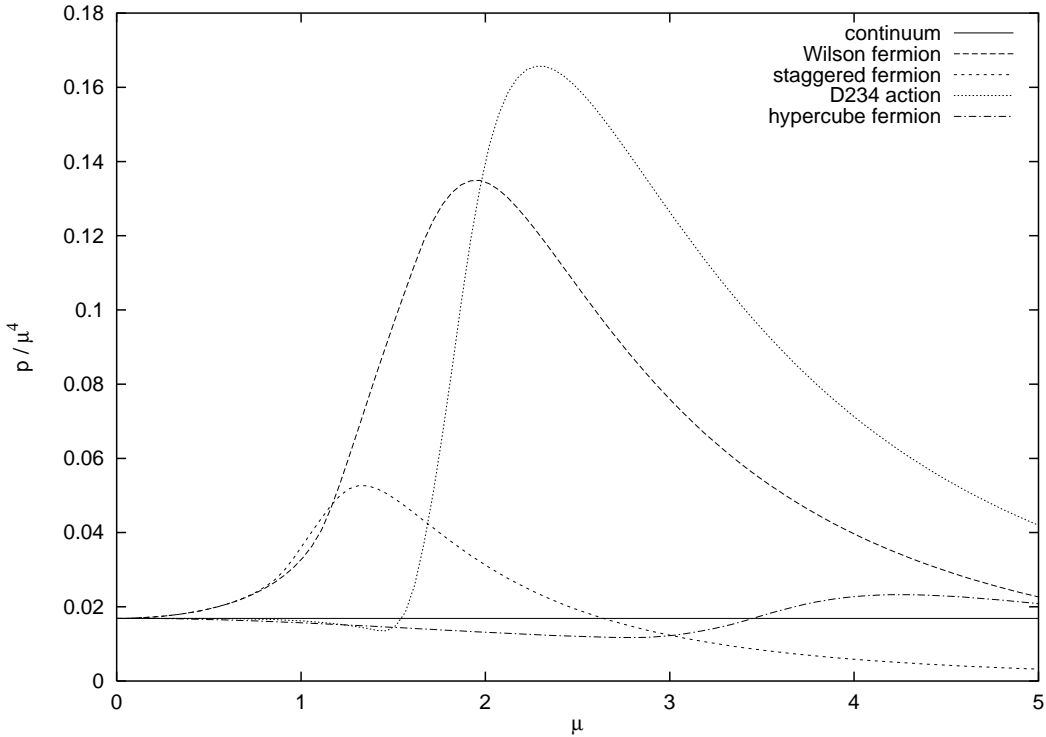}
\caption{\it The ratio $p/\mu^4$ as a function of the chemical
potential $\mu$ in lattice units. The solid line is the continuum
value $1/6 \pi^2$. The dashed lines corresponds to the Wilson action 
$(r=1)$, the staggered fermion action, the D234 action and the 
truncated perfect action (``hypercube fermion'').}
\end{figure}

Figure 2 illustrates the behavior of the baryon number
density $n_{B}$ divided by $\mu^3$. Again we compare the
Wilson fermion, the staggered fermion, the D234 action and
the truncated perfect action.
We see that the behavior of our second
scaling quantity, $n_{B}/\mu^{3}$, is qualitatively very similar
to $p/\mu^{4}$.
We summarize the outcome of our comparison again:
the Wilson and staggered fermion are plagued by large artifacts 
even at small $\mu$, 
the D234 action is good on fine lattices but disastrous
on very coarse ones, and the ``hypercube fermion''
stays by far closest to the continuum value. 
As a further example, the baryon number susceptibility
$\chi_{B}= \partial n_{B}/\partial \mu$ leads to a scaling
quantity $\chi_{B}/\mu^{2}$, which again behaves similarly
for the lattice fermions considered above. This is also in
qualitative agreement with comparisons done for other quantities 
before: in Ref. \cite{LAT96} we compared another thermodynamic
scaling ratio, $p/T^{4}$ at $\mu =0$, as well as the dispersion 
relations, and we found in both cases qualitatively the same
behavior as in Figures 1 and 2.

\begin{figure}[hbt]
\def\fpsangle{270}
\epsfxsize=100mm
\fpsbox{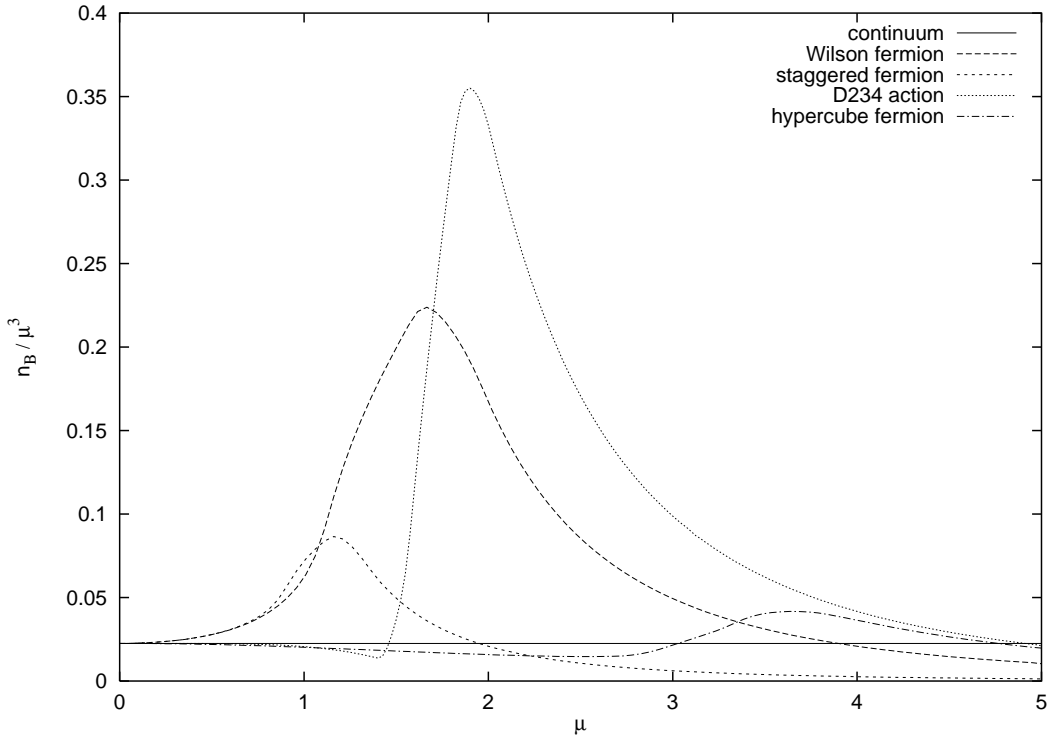}
\caption{\it The ratio $n_{B}/\mu^{3}$ as a function of the chemical
potential $\mu$ in lattice units. The solid line is the continuum
result $2/9 \pi^2$. The dashed lines corresponds to the Wilson action
$(r=1)$, the staggered fermion action, the D234 action and the truncated
perfect action (``hypercube fermion'').}
\end{figure}


Figure~3 illustrates the saturation effect at large chemical potential
for various lattice actions.
At large $\mu$, most lattice actions lead to an asymptotic
fermion number density $n_{f}=32$. That number corresponds to an 
occupation of each lattice point by 16 flavors with spin up and down.
This also includes the doubler fermions, which are not suppressed
any longer at sufficiently large $\mu$.
The Wilson action with Wilson parameter $r=1$ is special, 
because in that case the time-like doublers are completely 
removed from the spectrum (their mass diverges) \cite{SW}.
As a consequence, the asymptotic value for $n_{f}$ is reduced
to 16. For the staggered fermions the doub-ling is reduced by
construction, hence their saturation level is only 8.
It should be noted that a non-truncated perfect action does
not suffer from any saturation artifacts;
it behaves like the continuum action.
This is due to the
presence of couplings over infinite distances in Euclidean
time, even though these couplings are exponentially small.
In general, the maximal saturation level is proportional
to the largest time separation with non-zero couplings.
\footnote{This provides an argument that any (exactly)
perfect action (in more than one dimension) must couple
over infinite distances.}
Since the hypercube fermion has been truncated at time
separation 1, it saturates at $n_{f}=32$. From this point of view,
the D234 action could, in principle, reach $n_{f}=64$
because it couples over time separation 2.
However, similar to the effect described above for the Wilson action
with $r=1$, the actual saturation level is only half of that.
\footnote{What happens technically is that the apparently
dominant term at $\mu \to \infty$ cancels for the Wilson fermion
at $r=1$ and for the D234 action.}


Figure 3 further reveals that, as $\mu$ grows, the saturation 
occurs soon for the $r=1$ Wilson fermion, and at about the same point for
the D234 fermion. Staggered fermions saturate at even smaller $\mu$.
The saturation can be delayed, however, for Wilson fermions with
$r$ close to (but not equal to) 1, which generates a very heavy class
of doublers.
Then $n_{f}$ reaches a first plateau at 16, performs another jump 
as $\mu$ catches up with the
heaviest doubler mass, and finally saturates at $n_{f}=32$.
Also for the hypercube fermion the saturation is significantly 
delayed, and again we recognize a jump as $\mu$ exceeds
the heaviest doubler mass, so that the final saturation sets in.
And in this case, in contrast to the Wilson fermion with $r\approx 1$,
the delayed saturation goes along with an improved behavior at
small $\mu$.

\begin{figure}[hbt]
\def\fpsangle{270}
\epsfxsize=100mm
\fpsbox{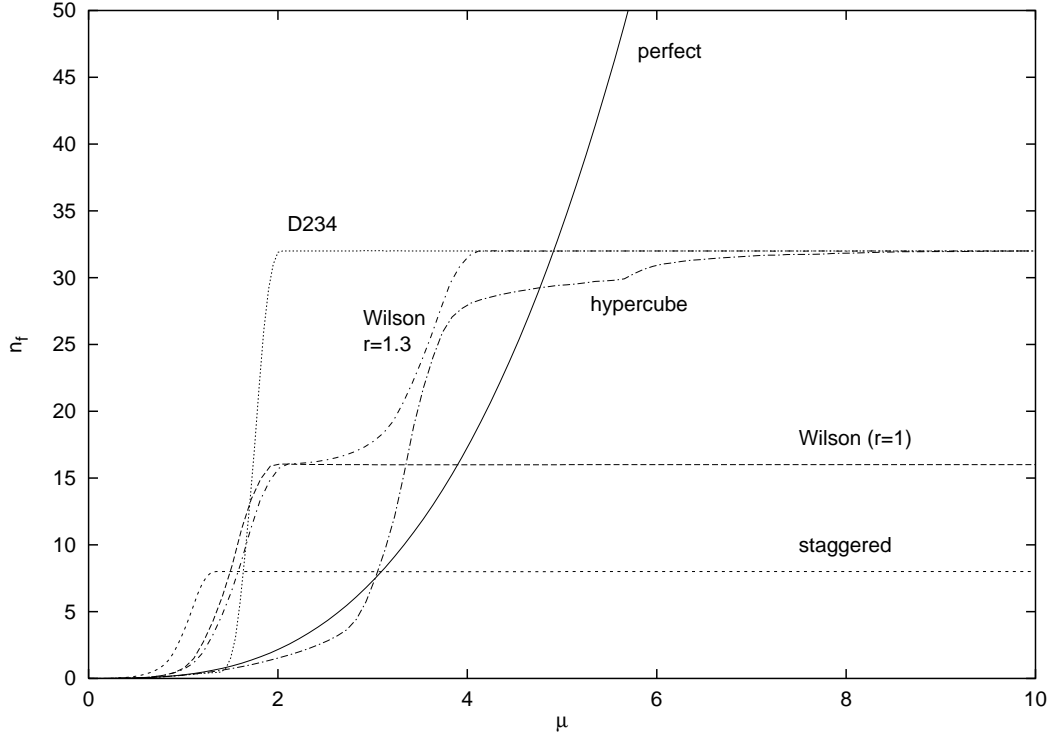}
\caption{\it The fermion number density $n_{f}$ as a function
of $\mu$ for various lattice fermions. The Wilson action with 
$r\neq 1$, the D234 action and the hypercube fermion saturate 
at $n_{f}=32$, whereas the Wilson fermion with $r=1$ (the staggered
fermion) only reaches $n_{f}=16$ ($n_{f}=8$).
Compared to the usual fermion actions, the
sa-turation is delayed significantly for the hypercube fermion.
For the perfect fermion action, $n_{f}$ is not bounded.}
\end{figure}

We have shown how to {\em incorporate a chemical
potential in a perfect lattice action} for full QCD.
In an application
to free fermions we observed a {\em strongly improved
scaling behavior} at finite $\mu$, even after truncation.
We further confirm the scenario that the unphysical saturation effect
is tamed 
for a (truncated) perfect action.
It is straightforward
to extend this study to more general cases, involving
for instance a finite quark mass
\footnote{A large fermion mass could be of
interest along the lines of the static quark method \cite{MILC}.}, 
a finite temperature or an anisotropic lattice.
The improvement permits lattice simulations for QCD at finite
baryon density to be performed on much coarser
lattices than it is the case for a standard
lattice action. Although there is also the major sign
problem due to the complex action,
such an improvement is important.

\ \\

Part of this work has been done at the University
of Erlangen. We thank for its hospitality.
We also thank M.-P. Lombardo
for reading the manuscript and for very helpful comments,
and F. Karsch for useful discussions.
Finally, we have benefited from our collaboration with
R. Brower, S. Chandrasekharan and K. Orginos.


\begin{thebibliography}{20}

\bibitem{HasKar} P. Hasenfratz and F. Karsch, Phys. Lett.
B 125 (1983) 308.\\
J. Kogut, H. Matsuoka, M. Stone, H. Wyld, S. Shenker, J. Shigemitsu
and D. Sinclair, Nucl. Phys. B225 [FS] (1983) 93.\\
N. Bili\'{c} and R. Gavai, Z. Phys. C23 (1984) 77.\\
R. Gavai, Phys. Rev. D32 (1985) 519.\\
The issue is reviewed in \\
I. Montvay and G. M\"{u}nster,
``Quantum Field Theory on the Lattice'', Cambridge Monographs, 1994.


\bibitem{Barb90} For an overview, see
I. Barbour, Nucl. Phys. B (Proc. Suppl.) 26 (1992) 22,
and references therein.

\bibitem{Japan} I. Barbour, S. Morrison, E. Klepfish, J. Kogut
and M.-P. Lombardo, Nucl. Phys. B (Proc. Suppl.) 60A (1998) 220.

\bibitem{barb} I. Barbour, N. Behilil, E. Dagotto, F. Karsch, A. Moreo,
M. Stone and H. Wyld, Nucl. Phys. B275 [FS17] (1986) 296.


\bibitem{Gibbs} P. Gibbs, Phys. Lett. B182 (1986) 369.


\bibitem{Glas} I. Barbour, S. Morrison, E. Klepfish, J. Kogut and M.-P.
Lombardo, Phys. Rev. D56 (1997) 7063.
 


\bibitem{Vink} J. Vink, Nucl. Phys. B323 (1989) 399.

\bibitem{Has94}
P. Hasenfratz and F. Niedermayer, Nucl. Phys. B414 (1994) 785.

\bibitem{DeG95}
T. DeGrand, A. Hasenfratz, P. Hasenfratz and F. Niedermayer, 
Nucl. Phys. B454 (1995) 587; 615.

\bibitem{Bie96} W. Bietenholz and U.-J. Wiese,
Nucl. Phys. B464 (1996) 319.

\bibitem{LAT96} W. Bietenholz, R. Brower, S. Chandrasekharan
and U.-J. Wiese, Nucl. Phys. B (Proc. Suppl.) 53 (1997) 921.

\bibitem{MaSto} R. Gavai and A. Ostendorf, Phys. Lett. 
B132 (1983) 137. \\
H. Matsuoka and M. Stone, Phys. Lett. B136 (1984) 204.

\bibitem{O3} P. Hasenfratz and F. Niedermayer, hep-lat/9706002.

\bibitem{Wie93} P. Ginsparg and K. Wilson, Phys. Rev. D25 (1982) 2649.\\
U.-J. Wiese, Phys. Lett. B315 (1993) 417.

\bibitem{Schwing} W. Bietenholz and U.-J. Wiese, Nucl. Phys. B
(Proc. Suppl.) 47 (1996) 575.

\bibitem{D234} M. Alford, T. Klassen and G.P. Lepage,
Nucl. Phys. B496 (1997) 377. 

\bibitem{FN} F. Niedermayer, Nucl. Phys. B (Proc. Suppl.) 53 (1997) 56.

\bibitem{SW} B. Sheikholeslami and R. Wohlert, Nucl. Phys. B259
(1985) 572.

\bibitem{MILC} T. Blum, J. Hetrick and D. Toussaint, Phys. Rev. Lett.
76 (1996) 1019.


\end{thebibliography}
\end{document}